\documentclass[%
reprint,
superscriptaddress,
 amsmath,amssymb,
 aps,
prb,
]{revtex4-1}

\usepackage{graphicx}
\usepackage{dcolumn}
\usepackage{bm}
\usepackage{hyperref}

\begin{document}
\title{Lattice and orbital fluctuations in TiPO$_4$}
\author{Dirk Wulferding} \affiliation{Institute for Condensed Matter
Physics, Technical University of Braunschweig, D-38106 Braunschweig, Germany}
\affiliation{Center for Artificial Low Dimensional Electronic Systems, Institute for Basic Science and Department of Physics, POSTECH, Pohang, 790-784, Korea}

\author{Angela M\"{o}ller} \affiliation{Department of Chemistry and Texas Center for Superconductivity, University Houston, Houston, Texas 77204-5003, United States}

\author{Kwang-Yong Choi} \affiliation{Department of Physics, Chung-Ang University, 221
Huksuk-Dong, Dongjak-Gu, Seoul 156-756, Republic of Korea}

\author{Yurii G. Pashkevich} \affiliation{A. A. Galkin Donetsk Phystech of the National Academy of Sciences of Ukraine, Donetsk 83114, Ukraine} \affiliation{ Donetsk National University, Donetsk 83001, Ukraine}

\author{Roman Yu. Babkin} \affiliation{A. A. Galkin Donetsk Phystech of the National Academy of Sciences of Ukraine, Donetsk 83114, Ukraine}

\author{Karina V. Lamonova} \affiliation{A. A. Galkin Donetsk Phystech of the National Academy of Sciences of Ukraine, Donetsk 83114, Ukraine} \affiliation{ Donetsk National University, Donetsk 83001, Ukraine}

\author{Peter Lemmens} \affiliation{Institute for Condensed Matter
Physics, Technical University of Braunschweig, D-38106 Braunschweig, Germany}

\author{Joseph M. Law}\affiliation{Dresden High Magnetic Field Laboratory, Helmholtz-Zentrum Dresden-Rossendorf, D-01314 Dresden, Germany}

\author{Reinhard K. Kremer} \affiliation{Max-Planck-Institut f\"{u}r Festk\"{o}rperforschung, Heisenbergstrasse 1, D-70569 Stuttgart, Germany}

\author{Robert Glaum} \affiliation{Institut f\"{u}r Anorganische Chemie, Universit\"{a}t Bonn, D-53121 Bonn, Germany}
\date{\today}

\begin{abstract}
In the $s=1/2$ antiferromagnetic spin chain material TiPO$_4$ the formation of a spin gap takes place in a two step process with two characteristic temperatures, $T^*= 111$ K and $T_{SP} = 74$ K. We observe an unusual lattice dynamics over a large temperature regime as well as evidence for an orbital instability preceding the spin-Peierls transition. We relate different intrachain exchange interactions of the high temperature compared to the spin-Peierls phase to a modification of the orbital ordering pattern. In particular, our observation of a high energy excitation of mixed electronic and lattice origin suggests an exotic dimerization process different from other spin-Peierls materials.

\begin{description}

\item[PACS numbers]{75.10.Pq, 75.25.Dk, 75.30.Kz}

\end{description}
\end{abstract}

\maketitle

\section{Introduction}

Quantum cooperative phenomena involving an interplay of electronic and lattice degrees of freedom are central issues in condensed matter physics.~\cite{Dagotto-Lee} Spin chains or coupled chain materials with considerable spin lattice interaction are a good basis for a spectroscopic study of such effects.~\cite{Lemmens, choi-bavs3} Instead of long-range magnetic order the magnetoelastic coupling can induce a cooperative change of the lattice and of the magnetic state by lowering the total energy of the system.~\cite{Bray} The so-called spin-Peierls (SP) transition leads to the formation of local magnetic singlets with a spin gap $\Delta$ as a consequence of alternating structural distortions between neighboring spins.

Experimental fingerprints of this transition are usually the appearance of comparably weak superstructure reflections in structural investigation methods due to a doubling of the unit cell, as well as a drop of the magnetic susceptibility due to the formation of a singlet ground state with a gap $\Delta$ to energetically higher triplet excitations. A crossover from well defined singlet and triplet bound states of dimerized phases to spinon excitations in the homogeneous state can be observed as a function of temperature or competing magnetic interactions. Such effects have been nicely demonstrated in CuGeO$_3$, the first inorganic SP system.~\cite{Hase} Later, TiO\textit{X} (\textit{X} = Br, Cl) has been found to represent a more complex case of dimer formation.~\cite{Seidel,Caimi,Shaz} This is due to the existence of an extended intermediate fluctuation regime for $T > T_{SP}$, leading to a decoupling of $\Delta$ from $T_{SP}$ with a ratio of $2 \Delta /T_{SP} \gg 3.5$, i.e. far beyond a BCS-like, mean-field model. For TiOCl the fluctuation regime has been attributed to comparably large, frustrated interchain interactions.~\cite{ruckamp-05} However, speculations about the role of orbital excitations in destabilizing the lattice dimerization and possible high temperature superconductivity have been made.~\cite{Seidel, lemmens-tiocl, kuntscher06}

Recently, Law {\it et al.},~\cite{law-11} reported a SP transition at $T_{SP}=74$ K in the antiferromagnetic quasi 1D $s=1/2$ system TiPO$_4$. TiPO$_4$ contains TiO$_6$ octahedra forming uniform chains along the crystallographic $c$ axis via shared edges (space group $Cmcm$). Measurements of the magnetic susceptibility reveal a dominating antiferromagnetic exchange interaction of $J = 965$ K between the $s=1/2$ Ti$^{3+}$ ions ($3d^1$).~\cite{law-11} The next-nearest neighbor exchange interaction $J_{nnn}=1.4$ K and an interchain coupling of $J_{ic}=20$ K are much smaller. Hence, a one-dimensional picture describes this system to a certain extent. Similar to TiO\textit{X}, two successive phase transitions were found from magnetic susceptibility and heat capacity measurements at $T^*=111$ K and $T_{SP}=74$ K. While the latter is due to a SP-like instability, the first one signals the transition from a uniform to an incommensurate dimerized spin chain.~\cite{law-11} From a fit to the magnetic susceptibility data, a spin gap of $\Delta = 830\,(425)$ K is estimated for the dimerized (incommensurate) phase.~\cite{law-thesis} Nonetheless, neutron powder diffraction studies of the crystal structure reveal no apparent evidence for a structural phase transition as function of temperature.~\cite{law-thesis, glaum-neutron} This highlights an apparent contradiction between the subtle nature of structural distortions and the relatively high SP transition temperature, which implies a strong spin-lattice interaction. Similarly, a strong deviation of the ratio $2 \Delta /T_{SP}$ from the BCS-like value points toward other non-magnetoelastic contributions to the SP mechanism.

Here, we report results of inelastic light scattering (Raman scattering) in TiPO$_4$, showing pronounced anomalies in phonon as well as magnetic scattering. From magnetic Raman scattering we can extract an antiferromagnetic intrachain coupling of $J = 620$ K in the SP phase, which is considerably lower than the value of the high temperature phase. We suggest that a first-order SP phase transition is accompanied by a Ti$^{3+}$ orbital reorientation. This modifies the magnitude of superexchange in accordance with the Goodenough-Kanamori rules. Semi-empirical crystal field calculations support these conclusions. The role of the orbitals in the dimerization process can account for an enhanced ratio $2\Delta / T_{SP}$, far beyond the mean-field BCS value. We consider two basic models of the SP phase with two possible mutual locations of dimerized Ti-Ti pairs on the neighboring chains: $i$) in-phase arrangement of adjacent dimerized Ti-chains with $Cm2m$ space group symmetry ($Amm2$ in standard setting); $ii$) out-of phase arrangement of adjacent dimerized Ti-chains with $Pmnm$ space group symmetry ($Pmmn$ in standard setting). We find that both these model space groups ($Pmnm$ and $Cm2m$) can describe the multiplication of phonon modes in the SP phase. In particular, the $Cm2m$ structure has the same translation symmetry as the high temperature phase. Our Raman studies provide evidence for the underlying dimerization process. However, one cannot resolve all further details of the low temperature structure of the SP phase based solely on Raman spectroscopic data. Thus, further synchrotron diffraction studies are needed.

\section{Experimental details}

Single crystals of TiPO$_4$ were grown by chemical vapor transport as described previously.~\cite{glaum-92} Typical crystal sizes are about $0.5 \times 0.5 \times 1$ mm$^3$. Sample characterization has been performed using X-Ray diffraction, magnetic susceptibility, nuclear magnetic resonance (NMR), electron spin resonance (ESR), and specific heat.~\cite{law-11, law-thesis}

Raman scattering experiments were performed in quasi-backscattering geometry, using a  $\lambda = 532$ nm solid state laser. The laser power was set to 5 mW with a spot diameter of approximately 100 $\mu$m to avoid heating effects. All measurements were carried out in an evacuated cryostat in the temperature range from 10 K to 295 K. The spectra were collected via a triple spectrometer (Dilor-XY-500) with a liquid nitrogen cooled CCD (Horiba Jobin Yvon, Spectrum One CCD-3000V).

\section{Experimental Results}

At high temperatures (i.e. above $T^*$), the factor group analysis for the orthorhombic space group $Cmcm$ (No. 63) with occupied Wyckoff positions for Ti-$(4a)$, P-$(4c)$, OI-$(8g)$, OII-$(8f)$ yields 15 Raman-active phonon modes: $\Gamma_{Raman} = 5 \cdot $A$_{1g} + 4 \cdot $B$_{1g} + 2 \cdot $B$_{2g} + 4 \cdot $B$_{3g}$. The corresponding Raman tensors are given by:

\begin{center}

\begin{widetext} \mbox{A$_{g}$=$\begin{pmatrix} a & 0 & 0\\ 0 & b & 0\\ 0 & 0 & c\\
\end{pmatrix}$

, B$_{1g}$=$\begin{pmatrix} 0 & d & 0\\ d & 0 & 0\\ 0 & 0 & 0\\ \end{pmatrix}$

, B$_{2g}$=$\begin{pmatrix} 0 & 0 & e\\ 0 & 0 & 0\\ e & 0 & 0\\ \end{pmatrix}$

, B$_{3g}$=$\begin{pmatrix} 0 & 0 & 0\\ 0 & 0 & f\\ 0 & f & 0\\ \end{pmatrix}$}

\end{widetext}
\end{center}

Raman scattering experiments were performed in different light polarizations at room temperature ($T = 295$ K) and at 10 K to distinguish between these modes. The assignment of the phonon modes to their respective symmetry is given in Table I. At room temperature, we clearly observe 12 out of the 15 expected phonon modes, which is in good agreement with the factor group prediction. The discrepancy between the observed and expected modes can be due to a lack of phonon intensity of particular modes as well as due to their overlap with excitations of larger intensity.

All Raman measurements were performed within a crystallographic plane spanned by the $c$ axis and a second axis that lies within the $ab$ plane, i.e. canted by about 45$^\circ$ from both $a$ and $b$ axis. Due to this peculiar crystal geometry, it is not possible to distinguish between modes of B$_{2g}$ and B$_{3g}$ symmetry. As the single crystal is rather thin and semi-transparent, a part of the light is scrambled in polarization and thus the selection rules for phonon modes of different symmetry are mixed. Therefore, the symmetry of three phonon modes marked by an asterisk in Table I is too ambiguous for a clear assignment. In the following, the scattering geometry $cc$ denotes both incoming and scattered light polarized along the crystallographic $c$ axis, while in the $cu$ geometry the analyzer is removed and all scattered light is sampled. Thereby the scattering intensity is maximized.

\begin{table*} \caption{\label{tab:table1}Phonon frequencies in cm$^{-1}$ with their respective symmetry assignments for the high temperature $Cmcm$ and the low temperature $Pmnm$ ($Cm2m$) phase measured at $T = 295$ K and at 10 K. Asterisks denote modes of ambiguous symmetry.}
\begin{ruledtabular} \begin{tabular}{ ccccc }
 \multicolumn{2}{c}{$T = 295$ K}&\multicolumn{2}{c}{$T = 10$ K}&\\
 Phonon frequency&Symmetry assignment&Phonon frequency&Symmetry assignment&Comment\\ \hline
  & &132&A$_g$ (A$_1$)&strong dynamics\\
  & &166&A$_g$ (A$_1$)&strong dynamics\\
  175&*&176&*&\\
	& &182&A$_g$ (A$_1$)&\\
  & &202&B$_{2g}$ / B$_{3g}$ (B$_1$ / B$_2$)&\\
  & &230&B$_{2g}$ / B$_{3g}$ (B$_1$ / B$_2$)&\\
  & &250&A$_g$ (A$_1$)&strong asymmetry\\
  276&B$_{1g}$&277&B$_{1g}$ (A$_2$)&\\
  & &280&B$_{1g}$ (A$_2$)&\\
  297&A$_g$&299&A$_g$ (A$_1$)&\\
  & &330&A$_g$ (A$_1$)&strong dynamics\\
  340&B$_{2g}$ / B$_{3g}$&342&B$_{2g}$ / B$_{3g}$ (B$_1$ / B$_2$)&\\
  & &366&A$_g$ (A$_1$)&\\
	& &400&B$_{2g}$ / B$_{3g}$ (B$_1$ / B$_2$)&\\
	& &429&B$_{2g}$ / B$_{3g}$ (B$_1$ / B$_2$)&\\
  451&A$_g$&452&A$_g$ (A$_1$)&\\
  486&B$_{2g}$ / B$_{3g}$&487&B$_{2g}$ / B$_{3g}$ (B$_1$ / B$_2$)&\\
  & &493&A$_g$ (A$_1$)&\\
  545&A$_g$&546&A$_g$ (A$_1$)&\\
  665&*&665&*&\\
	& &943&A$_g$ (A$_1$)&\\
  948&B$_{1g}$&949&B$_{1g}$ (A$_2$)&\\
  960&A$_g$&960&A$_g$ (A$_1$)&\\
  & &964&A$_g$ (A$_1$)&\\
  & &1041&A$_g$ (A$_1$)&\\
  1047&B$_{1g}$&1048&B$_{1g}$ (A$_2$)&\\
  & &1070&B$_{2g}$ / B$_{3g}$ (B$_1$ / B$_2$)&\\
  1138&*&1139&*&\\
\end{tabular} \end{ruledtabular} \end{table*}

\begin{figure}
\label{figure1}
\centering
\includegraphics[width=8cm]{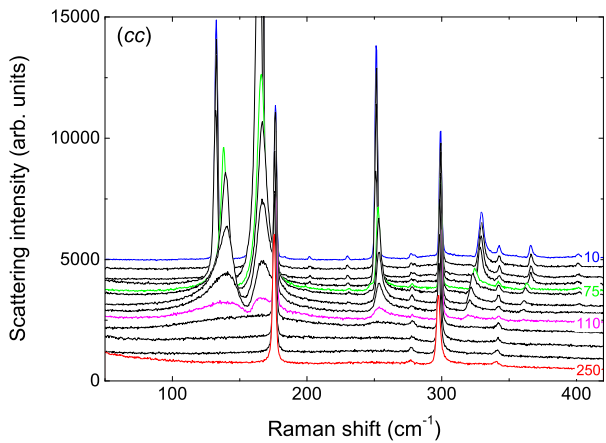}
\caption{(Color online) Raman spectra collected at $T$ = 10, 30, 50, 70, 75, 80, 90, 100, 110, 120, 150, 200, and 250 K in $cc$ polarization, i.e. along the chain direction. The depicted spectra range from 50 to 420 cm$^{-1}$. They are shifted in intensity for clarity.}
\end{figure}

In Fig. 1 the low frequency regime of the Raman spectra is shown for different temperatures ranging from 10 to 250 K. Here, four modes at 132, 166, 250, and 330 cm$^{-1}$ show a pronounced temperature dependence. To trace the temperature development of the phonon energies, we plot the frequencies as a function of temperature for each phonon mode observed in $cc$ polarization in Fig. 2. We note two static regimes for temperatures above $T^*$ and temperatures below $T_{SP}$, signalling a stable crystal lattice. In contrast, in the intermediate temperature regime below $T^*$ and above $T_{SP}$ several new phonon modes occur, denoted by open red circles. These modes exhibit a strong dynamical behavior, evidenced by anomalously large shifts in frequency of up to 16 cm$^{-1}$. In the lower panel of Fig. 2 we zoom into the spectral regime of PO$_4$ intrinsic vibrations, obtained in $cu$ polarization, to detail the occurrence of new phonon modes due to the dimerization process.

\begin{figure}
\label{figure2}
\centering
\includegraphics[width=8cm]{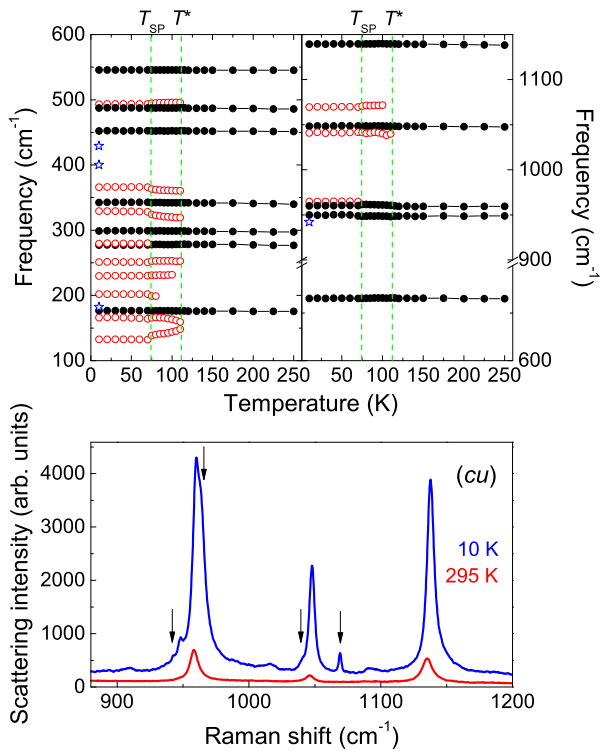}
\caption{(Color online) Upper panel: Temperature dependence of the phonon frequencies observed in $cc$ polarization. Changes at the two characteristic temperatures $T^*$ and $T_{SP}$ are clearly visible together with the occurrence of new phonon modes (empty red circles). The blue stars denote phonon modes of low intensity, observed only in $cu$ polarization. Lower panel: Comparison of the intrinsic PO$_4$ phonon modes at $T = 295$ and 10 K obtained in $cu$ polarization. The arrows mark new phonon modes in the low temperature phase.}
\end{figure}

At low temperatures, the overall number of observed phonon modes increases to 28, clearly signalling a lowered crystal symmetry. First principle density functional theory calculations were carried out in order to predict the low temperature crystal structure of TiPO$_4$.~\cite{law-thesis} It was found that an energy minimum is reached for the $Pmnm$ (No. 59) space group with occupied Wyckoff positions for Ti-$(4e)$, PI-$(2a)$, PII-$2b$, OI-$(4e)$, OII-$(4e)$, OIII-$(4f)$, OIV-$(4f)$. This yields a total of 36 Raman active phonon modes ($\Gamma_{Raman}(Pmnm) = 12 \cdot$ A$_g + 5 \cdot$ B$_{1g} + 9 \cdot$ B$_{2g} + 10 \cdot$ B$_{3g}$). In this case additional phonon modes are induced from the high temperature Brillouin zone boundary at the Y-point, i.e. at $\textbf{k}=(0,2\pi/b,0)$. A second possible low temperature space group with a slightly higher calculated ground state energy is $Cm2m$ (No. 38) with occupied Wyckoff positions for Ti-$(4c)$, PI-$(2a)$, PII-$(2b)$, OI-$(4d)$, OII-$(4e)$, OIII-$(4c)$, OIV-$(4c)$. This yields 33 Raman active optical phonon modes ($\Gamma_{Raman}(Cm2m) = 11 \cdot$ A$_1 + 5 \cdot$ A$_2 + 9 \cdot$ B$_1 + 8 \cdot$ B$_2$).

\begin{figure}
\label{figure3}
\centering
\includegraphics[width=8cm]{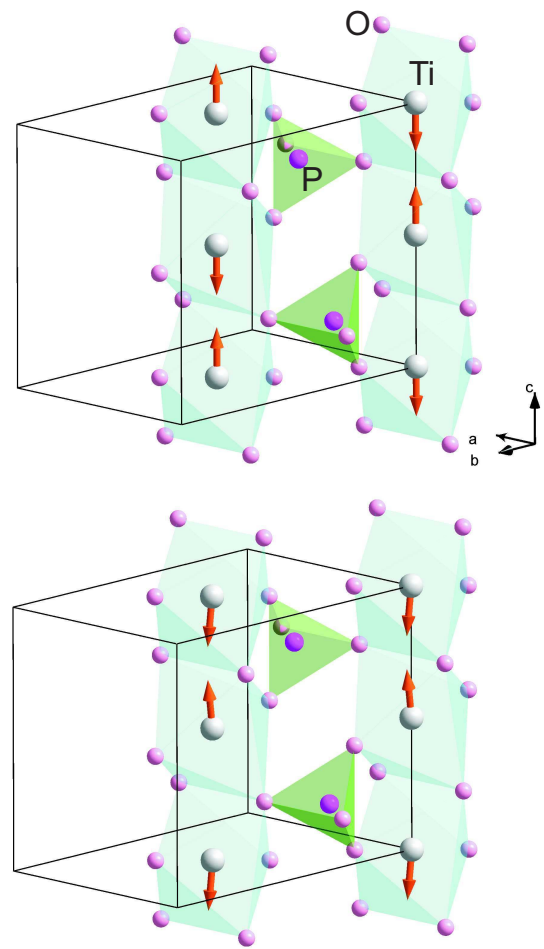}
\caption{(Color online) Two types of primary structures with Ti-dimerizing shifts which can serve as building blocks for more complex SP order. Upper panel: out-of-phase dimerization on neighboring chains. This structure is in accordance with the $Pmnm$ space group. Lower panel: in-phase dimerization on neighboring chains. This structure is in accordance with the $Cm2m$ space group.}
\end{figure}

In Fig. 3 we present the two possible scenarios for a SP distortion: the dimerization pattern in the upper panel refers to the out-of phase dimerization and leads to the $Pmnm$ space group. Here the glide plane is preserved but not the Bravais type of lattice ($c$-centered). The in-phase arrangement of adjacent dimerized chains can occur without a change of the Bravais type of lattice and is described by the $Cm2m$ space group (shown in the lower panel of Fig. 3). For this type of SP distortion the centered unit cell translation symmetry is preserved but the $c$-glide plane is now removed. Therefore, the $Cm2m$ structure deserves our special consideration. In the $Cm2m$ space group the rotation symmetry is lowered from $D_{2h}$ to $C_{2v}$ with a loss of inversion and the only twofold axis is directed along $b$. Therefore, all previously far-infrared (FIR) active phonon modes as well three A$_u$ silent modes become Raman active. In FIR experiments 11 phonon modes out of 15 FIR active ones have been observed at room temperature.~\cite{baran-89} A comparison with our data shows that six of them can be assigned to new ones which appear at low temperature. Meanwhile, the phonon frequencies of the remaining ones might be in the vicinity of high temperature Raman active modes with strong intensity. Note that for both $Pmnm$ and $Cm2m$ space groups the phosphorus ions occupy two different positions in accordance with NMR observations.~\cite{law-11}

The two considered model structures will act as building blocks for more complex dimerization processes that will eventually lead to larger units cells, but not necessarily require a doubling of the $c$ axis. In our low temperature data we observe 28 phonon modes. We can identify 14 modes that can be ascribed to A$_g$ (A$_1$) symmetry and 4 modes with B$_{1g}$ (A$_2$) symmetry. Our observations do not contradict the $Pmnm$ ($Cm2m$) space group model. However, there is a discrepancy of 8 (5) phonon modes as in total only 28 out of 36 (33) modes are observed. The missing modes are mostly of B$_{2g}$ (B$_1$) and B$_{3g}$ (B$_2$) symmetry. These modes could be masked by other modes of strong intensity. In order to ultimately distinguish between the two model space groups or even include more complicated mixtures of both scenarios at low temperatures, high resolution synchrotron measurements would be required.~\cite{LT-remark}

We can separate the phonon modes into three regimes. This also agrees with previous infrared study on TiPO$_4$,~\cite{baran-89} and recent lattice dynamic calculations.~\cite{Lopez-Moreno} Phonons above $\sim$ 500 cm$^{-1}$ are related to PO$_4$ modes, phonons occurring in the intermediate frequency range (400 -- 500 cm$^{-1}$) are related to the coupling of the building blocks PO$_4$ and TiO$_6$, and phonons below $\sim$ 400 cm$^{-1}$ are predominantly related to oxygen displacements along the chain direction originating from TiO$_6$ units. While the Ti ions themselves are not Raman active in the high temperature phase, they are the heaviest ions in TiPO$_4$ and therefore contribute to the low frequency part of the phonon spectra. The modulation of Ti-Ti distances and Ti-O-Ti angles are strongly affected by their vibrations.

\begin{figure}
\label{figure4}
\centering
\includegraphics[width=8cm]{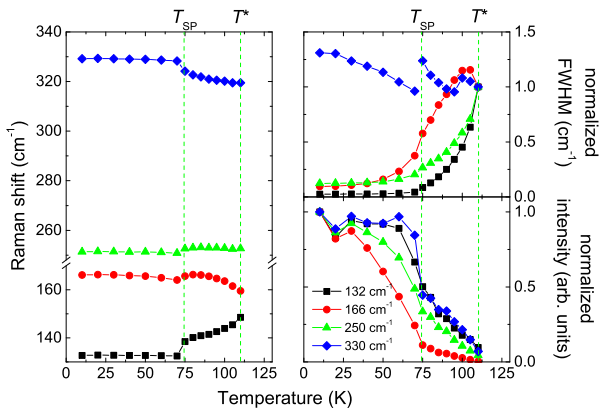}
\caption{(Color online) Temperature dependence of the frequency, normalized linewidth (full width at half maximum) and normalized intensity for the A$_g$ modes at 132 cm$^{-1}$ (black squares), 166 cm$^{-1}$ (red circles), 250 cm$^{-1}$ (green triangles), and 330 cm$^{-1}$ (blue diamonds).}
\end{figure}

To detail the crystal's lattice dynamics, we plot the frequency, linewidth and intensity of four A$_g$ phonon modes with very strong dynamical behavior in Fig. 4. All four modes only appear for temperatures below $T^*$ and become static for temperatures below $T_{SP}$. Their energies are below 400 cm$^{-1}$ and they display both softening and hardening with temperature. We follow from this unusual temperature dependence that modes in this frequency range display a strong interplay between lattice degrees of freedom and the magnetic subsystem.

In the left panel of Fig. 5 we plot spectra up to 700 cm$^{-1}$ at four different temperatures ranging from 10 to 250 K to highlight the temperature evolution of the magnetic scattering. The inset traces its integrated intensity as a function of temperature. In the right panel we extend the spectral window up to 3000 cm$^{-1}$ for the spectra at 10 K and at room temperature (295 K), showing the complete spectral range of the magnetic continuum. Note, that we repeated this measurement with several excitation lines from $\lambda = 488$ to 647 nm (not shown) to rule out fluorescence as the origin of this enhanced spectral weight.

\begin{figure*}
\label{figure5}
\centering
\includegraphics[width=15cm]{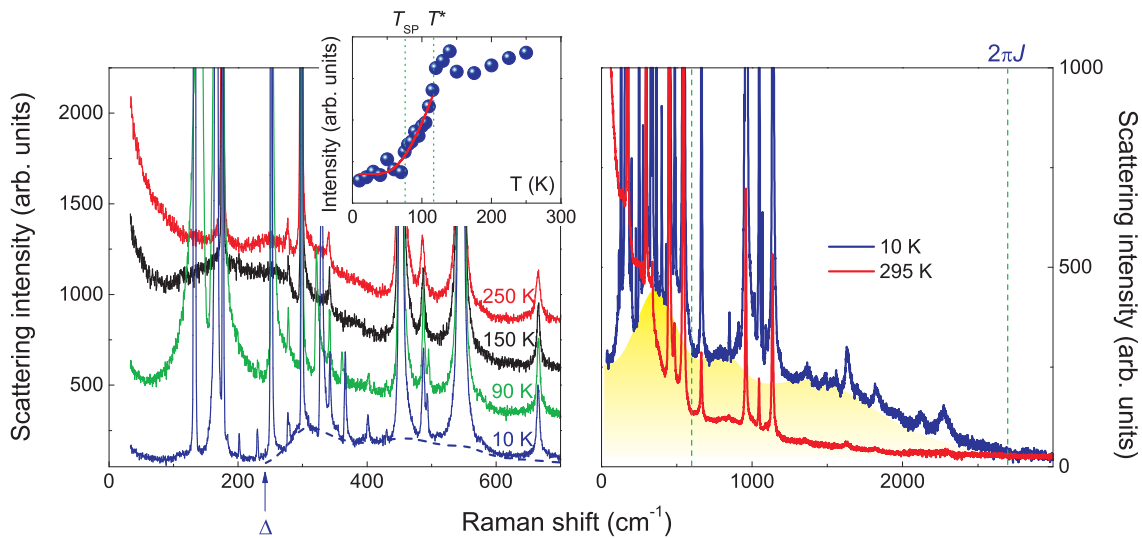}
\caption{(Color online) Left panel: Raman spectra obtained at 10, 90, 150, and 250 K. The dashed blue line denotes the magnetic scattering continuum at 10 K. The spectra are shifted in intensity for clarity. The inset depicts the integrated intensity of the continuum as a function of temperature. Right panel: Raman spectra obtained at base- and at room temperature. Shown is the complete spectral range of the magnetic scattering continuum. The dashed green lines mark the cut-off energies around 600 cm$^{-1}$ and 2700 cm$^{-1}$ ~for the pronounced spectral weight and the broad spinon-like continuum, respectively.}
\end{figure*}

In order to obtain information on the orbital order pattern for the different phases, we calculate the electronic energy levels using a modified crystal field approach. The results of these calculations are shown in Fig. 6, while in Fig. 7 we plot the resulting electron density distributions in the high temperature phase ($T > T^*$) and in the dimerized phase ($T < T_{SP}$).

\begin{figure}
\label{figure6}
\centering
\includegraphics[width=8cm]{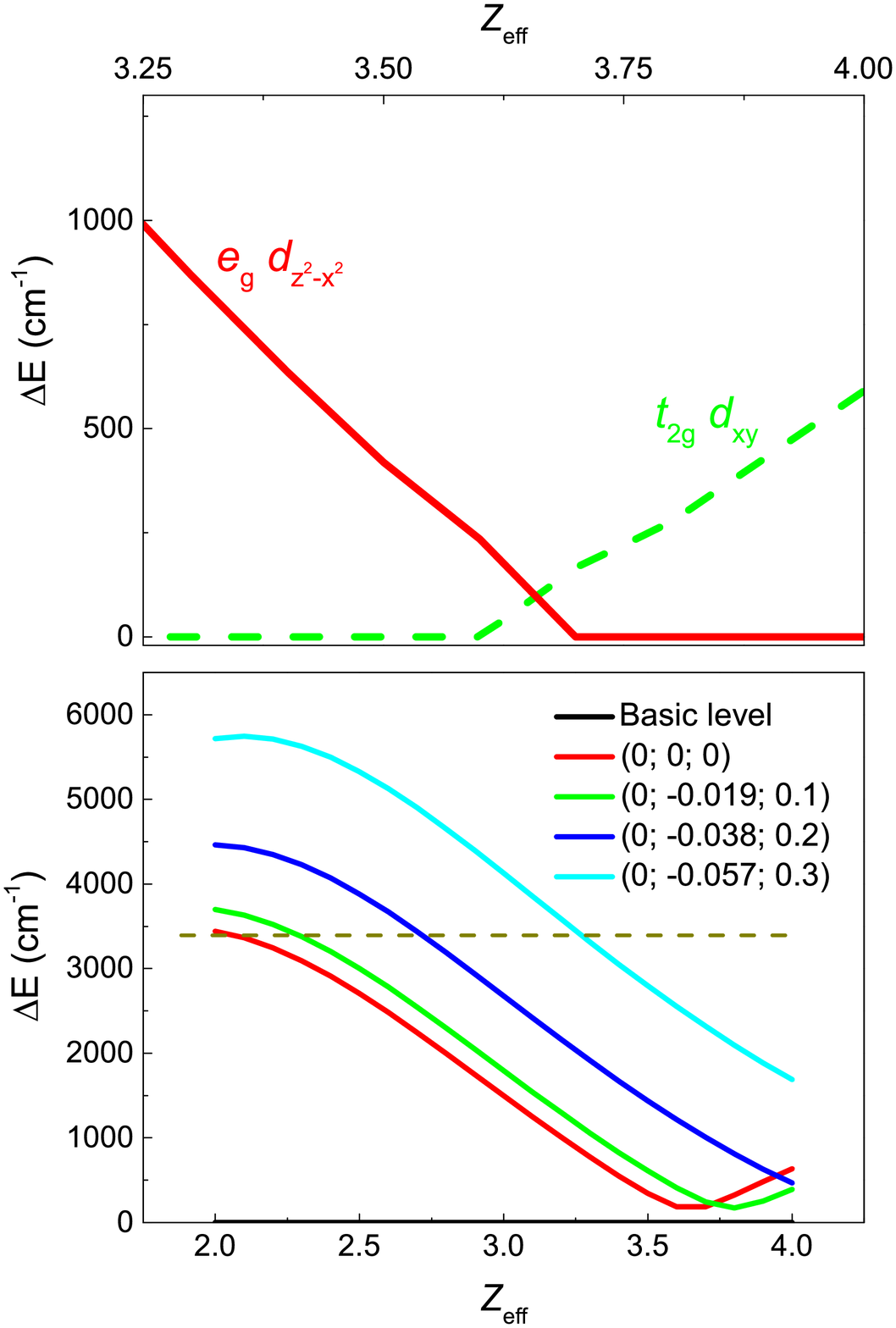}
\caption{(Color online) Upper panel: The ground state orbital cross-over and accidental degeneracy of the crystal field levels in TiPO$_4$ under lowering the Ti$^{3+}$ effective nuclear charge $Z_{eff}$ (by increasing covalency). The critical value of $Z_{eff}$ is $\bar{Z}_{eff}=3.62$. Lower panel: The disappearance of the ground state orbital crossover and onset of the $d_{xy}$ orbital ground state under the dimerization process is shown. The cartesian components of the Ti off-centering are given in brackets. The evolution of the first excited crystal field level is shown. The horizontal line marks the experimentally found value of the first excited crystal field level.}
\end{figure}

In addition to the phonons, a high energy mode appears at 3430 cm$^{-1}$ (shown in Fig. 8) with a sharp linewidth and a relatively strong intensity at low temperatures (i.e. below $T_{SP}$). With increasing the temperature above the SP transition its intensity quickly drops off, while the linewidth increases drastically. In addition, we observe a pronounced jump in energy around $T_{SP}$, as depicted in the right panel of Fig. 8. The energy of this mode is too large for phonon or magnetic scattering processes. Instead, its energy fits well to the titanium crystal field onsite excitations observed in TiOCl by resonant inelastic X-Ray scattering.~\cite{Glawion} A weaker maximum is observed at slightly higher energies and the separation from the main line corresponds to typical phonon energies. This is a characteristic feature of exciton-phonon multiparticle processes.

\begin{figure}
\label{figure7}
\centering
\includegraphics[width=8cm]{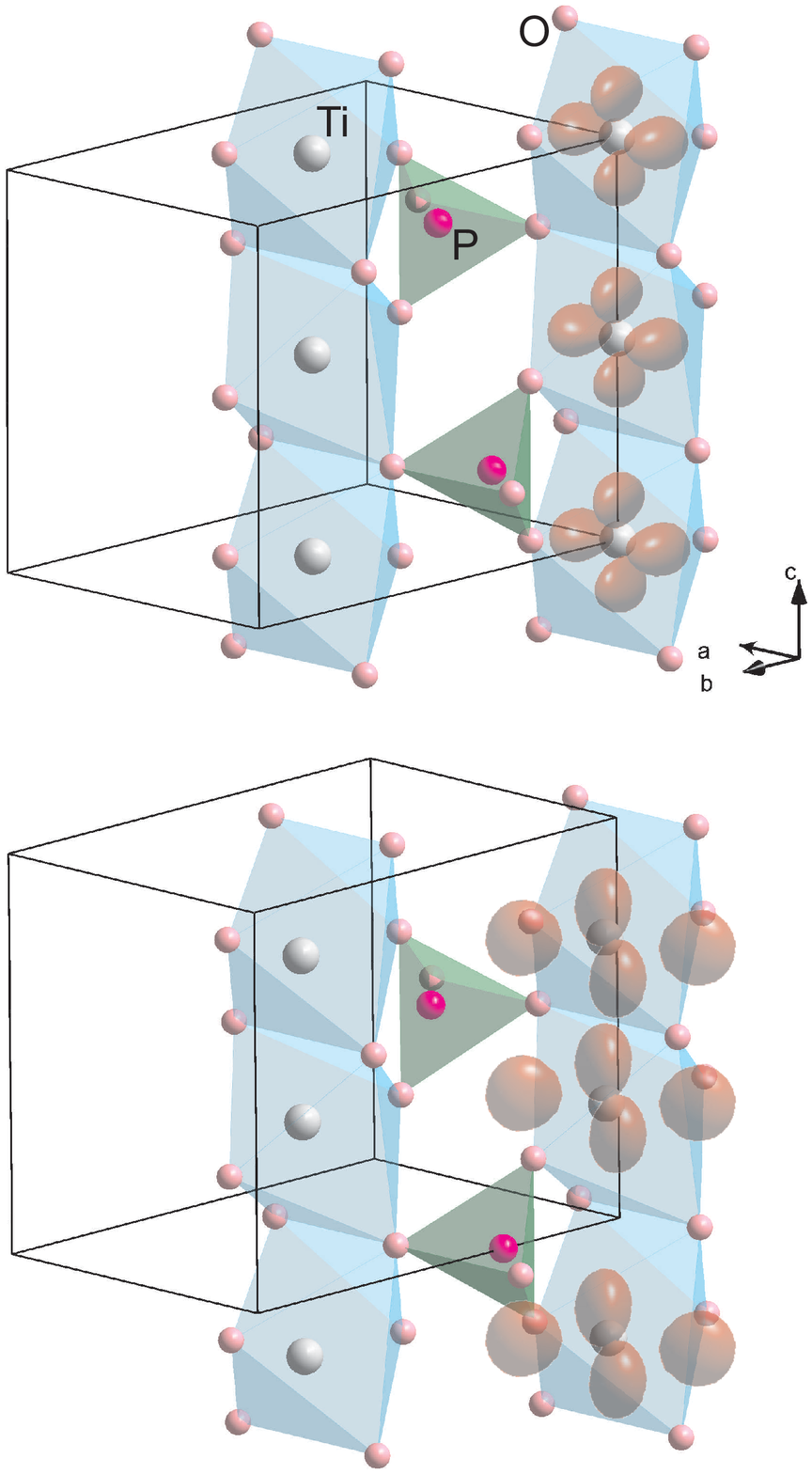}
\caption{(Color online) Upper panel: orbital order in the high-temperature phase $T > T^*$. Lower panel: orbital order in the SP phase for $T < T_{SP}$.}
\end{figure}

\begin{figure}
\label{figure8}
\centering
\includegraphics[width=8cm]{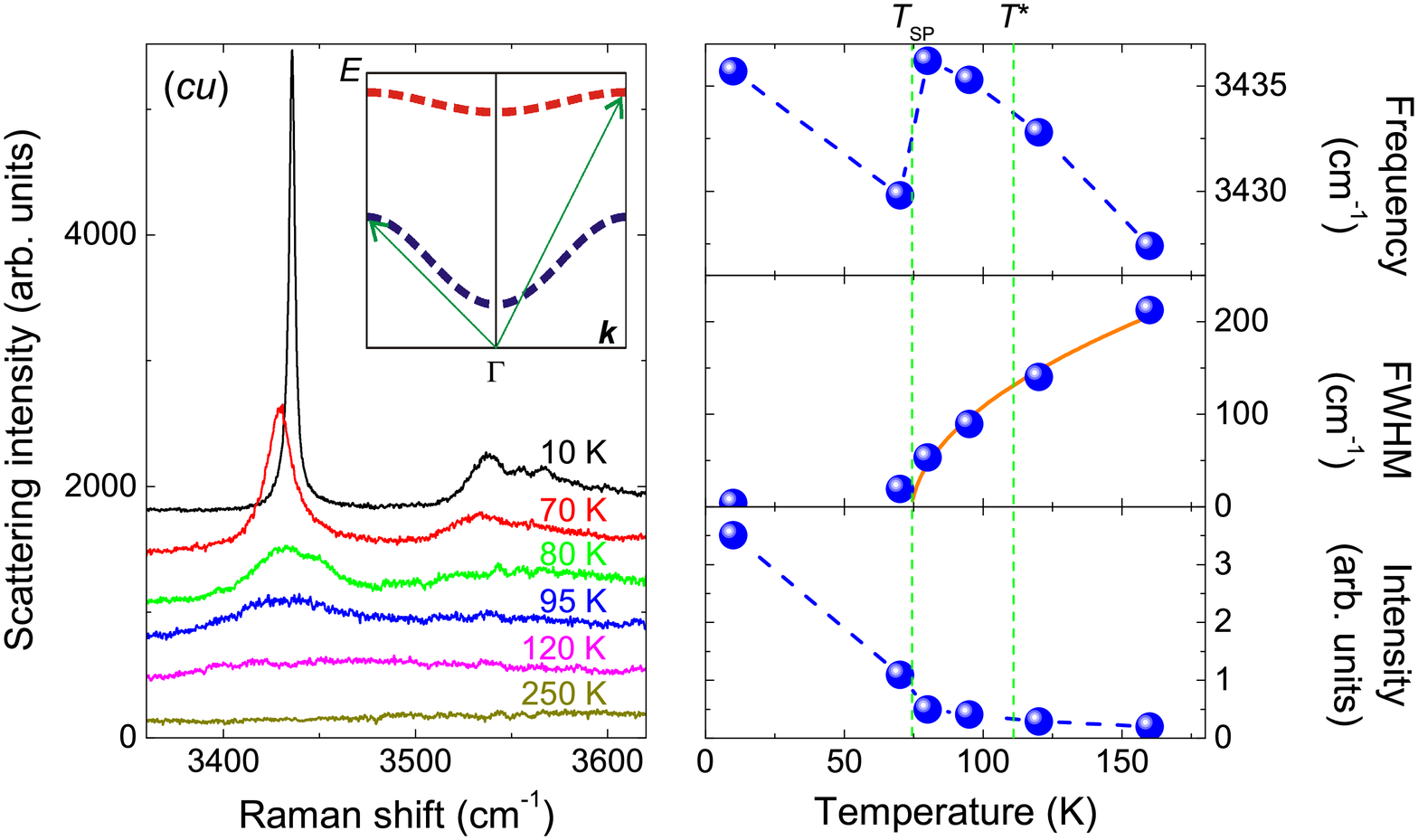}
\caption{(Color online) Left panel: Raman spectra of the high energy mode around 3430 cm$^{-1}$ obtained at different temperatures in $cu$ polarization. The inset sketches the scattering mechanism of the high frequency shoulders. Right panel: temperature dependence of the frequency, linewidth, and intensity of the high energy mode. The orange curve is a fit to the linewidth (see text for details).}
\end{figure}

\section{Discussion}

We will first focus on the phonon dynamics before moving on to the magnetic continuum and the high energy mixed excitations. In SP systems, spin-phonon coupling can account for strong lattice dynamics. To estimate the spin-phonon coupling in TiPO$_4$, we derive the spin-spin correlation function $\langle s_i,s_j \rangle_\chi$ from the magnetic susceptibility $\chi_m(T)$.~\cite{law-11} Considering only nearest neighbor interactions, the spin-spin correlation function can be written as~\cite{kant-spin-phonon}
\begin{equation}
\langle s_i,s_j \rangle_\chi = \frac{k_B T \chi_m(T)}{N_A g^2 \mu_B^2} - \frac{s(s+1)}{3}
\end{equation}
with $g=1.94$ and $s=1/2$. We can now estimate the spin-phonon coupling constant $\lambda$ by comparing the shift in phonon frequency to the spin-spin correlation function according to
\begin{equation}
\Delta \omega = \lambda \cdot \langle s_i,s_j \rangle_\chi
\end{equation}
This yields a value of the order of $\lambda_{132} \approx 100$ cm$^{-1}$ for the phonon mode at 132 cm$^{-1}$. The corresponding spin-phonon coupling constants of the other three phonon modes with strong dynamics at 166, 250, and 330 cm$^{-1}$ amount to $\lambda_{166} \approx -40$ cm$^{-1}$, $\lambda_{250} \approx 15$ cm$^{-1}$, and $\lambda_{330} \approx -80$ cm$^{-1}$. These values are larger than those in the SP system CuGeO$_3$, where spin-phonon coupling constants of $\lambda_{103} = -10$ cm$^{-1}$, $\lambda_{215} = 40$ cm$^{-1}$, $\lambda_{366} = -21$ cm$^{-1}$, and $\lambda_{812} = -8$ cm$^{-1}$ were found.~\cite{werner-99} This underlines the rather strong coupling for certain modes in TiPO$_4$. In particular, $\lambda_{132}$ is too large to be based on purely magnetoelastic coupling.

A proof of the Ti-Ti dimerization at low temperatures can be found in the high frequency phonon spectra (see lower panel of Fig. 2). The existence of at least two different phosphorous sites is an obligatory condition for the dimerization process as it involves different connectivities to Ti-pairs and linking to adjacent chains of in-phase and out-of phase Ti-chain arrangements, respectively. This should be seen in a change in the PO$_4$ vibrational spectra. In TiPO$_4$ the phonons above $\sim$ 900 cm$^{-1}$ are related solely to the PO$_4$ intrinsic bond stretching modes.~\cite{baran-89,Lopez-Moreno} Several new phonon modes occur in our Raman spectra at low temperatures in this frequency region (see lower panel of Fig. 2). This observation is in accordance with previous NMR studies.~\cite{law-11}

Focusing on the magnetic excitations in TiPO$_4$, we observe a temperature dependent scattering contribution (see Fig. 5, left panel) that is superimposed onto a broad continuum (see Fig. 5, right panel). At 10 K, one observes a background of magnetic origin with a clear onset energy of 250 cm$^{-1}$, highlighted by the dashed blue line in the left panel. Upon increasing temperature, this background shows distinct changes as the spectral weight increases and shifts towards lower energies.

In the inset of Fig. 5 we trace the temperature evolution of the continuum's integrated intensity. Upon the opening of the spin gap $\Delta$ the spectral weight of the magnetic scattering decreases exponentially with temperature. By applying an exponential fit function (red line) $\Gamma (T) = B + C \cdot e^{-\Delta / k_B T}$ to the data points below $T \approx$ 110 K, we obtain a gap size of $\Delta = (311 \pm 48)$ K. Note that for this analysis, only the spectral weight up to 600 cm$^{-1}$ ($\mathrel{\widehat{=}}$ 860 K) was taken into account as the influence of the temperature (10 -- 295 K) is most dramatic in this low energy regime. Here, we also note the asymmetric, Fano-like line shape of phonon modes in the low energy range, due to a coupling between the phonons and the continuum of magnetic excitations (see the left panel of Fig. 5).

A second, independent determination of the gap size can be achieved directly from the spectra, since a spin gap suppresses magnetic scattering below the gap energy. We observe the onset of the magnetic scattering at lowest temperatures at around 250 cm$^{-1}$. This corresponds to a gap of $\Delta \approx 360$ K, which is in good agreement with the gap size obtained from the exponential fit. Hence, the averaged spin gap value in TiPO$_4$ is $\Delta \approx 330$ K. This value is much smaller than the gap extracted from a fit to the magnetic susceptibility ($\Delta_{sus} = 830$ K,~\cite{law-thesis}). The discrepancy might be related to the considerable residual magnetic susceptibility in the gapped, low temperature phase.~\cite{law-11}

For a conventional SP mechanism, the ratio between spin gap and transition temperature is given by $2\Delta / T_{SP} = 3.5$, based on a mean field BCS-type relation.~\cite{regnault-96} In CuGeO$_3$, the ratio corresponds to $\sim 3.4$. Thus, its physics can be well described by a conventional SP mechanism. In contrast, in TiOCl the ratio amounts to 10 -- 15, which cannot be explained simply by a BCS-type SP mechanism. Instead, for this compound substantial frustrating interchain interactions ($J' \sim 0.2 \cdot J$)~\cite{lemmens-tiox} have been assigned to the observed fluctuations and the reduction of $T_{SP}$. For TiPO$_4$ with $2\Delta / T_{SP}=9$, such a scenario is improbable as the interchain interactions are rather small. This suggests that another source of fluctuations should be taken into account. Since orbital excitation energies are closer to the characteristic energies of the magnetic subsystem compared to other SP systems, orbital fluctuations might be responsible for the intermediate fluctuation regime and the reduction of $T_{SP}$.

Another difference between CuGeO$_3$ and TiPO$_4$ occurs in the dimerized phase at $E < \Delta$. While the magnetic scattering intensity is totally suppressed in the former system, a substantial finite spectral weight remains in TiPO$_4$ (see the enhanced low energy spectral weight at $T = 10$ K in the right panel of Fig. 5). In addition, the continuum of spinon excitations at high energies is considerably damped even at room temperature. This points to a complex nature of dimerization.

The effect of thermal fluctuations may lead to a cross-over in the dynamics of quantum systems (see right panel of Fig. 5). The high energy spectral weight has an upper limit around 2700 cm$^{-1}$ and is governed by spinon excitations of fermionic nature. In contrast, the low energy regime in the dimerized, low temperature phase consists of bosonic singlet to triplet excitations. With increasing temperature the transition from a dimerized to a uniform chain closes the spin gap wherein the low-energy bosonic excitations dampen in energy to $\omega \rightarrow 0$. Hence, this energy regime is replaced by a diffusive, quasielastic tail due to thermally induced fluctuations of the magnetic energy density.

This qualitative description fits also very well to the calculated and the experimentally observed magnetic Raman scattering in CuGeO$_3$.~\cite{loosdrecht-96} While the continuum for a two-particle scattering process in both CuGeO$_3$ and TiPO$_4$ should reach up to $4\pi J$  at the Brillouin zone boundary, thermal fluctuations at finite temperatures suppress scattering from the second half of the Brillouin zone ($\pi / 2 < k < \pi$).~\cite{loosdrecht-96} Therefore, a filtering function $f_k = \mathrm{cos}^2(k/2)$ is introduced for both CuGeO$_3$ and TiPO$_4$. This restricts the magnetic scattering spectra to an upper limit of $E_{cut-off} \approx 2\pi J$. From the observed cut-off energy a rough estimation of the magnetic exchange constant $J = 620$ K is extracted. Even taking quantum renormalization into account the esimated value of intrachain exchange in the SP phase strongly deviates from the reported value extracted from magnetic susceptibility in the high temperature phase ($J = 965$ K). Such a difference can be explained only under the assumption of orbital reordering during the dimerization process, which in accordance with Goodenough-Kanamori rules should change the magnitude of the intrachain exchange.

To study the possibility for unusual orbital ground state reorder in TiPO$_4$ we apply the modified crystal field approach (MCFA). This method is capable of calculating the electronic energy levels of an arbitrarily distorted coordination complex at arbitrary values of the ligand's charge with the spin-orbit interaction of the transition metal.~\cite{zhit-07, babkin-09, lamonova-11} The main variable in this approach is the effective nuclear charge $Z_{eff}$ of the metal ion. The variation of $Z_{eff}$ implicitly accounts for the variation of covalency of the ``metal-ligand'' bonds. For instance, as the temperature decreases, the ``metal-ligand'' bonds become more covalent and $Z_{eff}$ decreases. Off-centering of the metal ion, as it takes place during the dimerization process, also decreases $Z_{eff}$. In general, in the crystal field environment the effective nuclear charge can be reduced up to 20\% compared to its ``Slater'' value~\cite{Slater} of the free ion. In the case of free Ti$^{3+}$ ions, $Z_{eff}$ has a value of 4.

The results of the MCFA calculations are shown in the upper panel of Fig. 6. We used crystallographic data at 295 K and at 100 K from Ref.~\cite{law-thesis} We set the ligand's charge to -2 and the constant of the spin-orbit interaction for the free Ti$^{3+}$ ion to 120 cm$^{-1}$. In the upper panel of Fig. 6 we present the first excited crystal field level. We conclude that there is a critical value of $\bar{Z}_{eff} \approx 3.6$ at which the energy of the first excited crystal field level drops to zero and an accidental degeneracy of the Ti$^{3+}$ electronic levels occurs. Furthermore, for $Z_{eff} > \bar{Z}_{eff}$ the orbital ground state represents an unusual mixture of $e_{g}$-type $d$-wave functions, while for $Z_{eff} < \bar{Z}_{eff}$ the orbital ground state is mainly composed of $d_{xy}$-wave functions with $t_{2g}$-type symmetry.~\cite{remark}

The calculated electron density for $Z_{eff} > \bar{Z}_{eff}$ is shown in Fig. 7 (top). It can be roughly described by the $e_g$-type $d_{z^{2}-x^{2}}$-wave function. This symmetry of electron density with two pairs of lobes directed along the $c$- and the $a$-axes was determined by X-Ray single-crystal diffraction experiments in the high temperature phase at temperatures between 293 and 90 K.~\cite{law-11} Moreover, comparing electron density calculations for 295 K and 100 K we observe a migration of electron density from $a$-lobes into $c$-lobes with lowering temperature, in accordance with experiments.~\cite{law-11}

We model the dimerization process by off-centering the Ti$^{3+}$ ions in the octahedra along the crystallographic $c$-axis, i.e. along chain direction. The behavior of the first excited level for shifts of different magnitude ($\delta c = 0.1; 0.2; 0.3$ \text{\AA}) is shown in the lower panel of Fig. 6. The accidental degeneracy point disappears (i.e. $\bar{Z}_{eff}$=4) already at very modest shifts $\delta c=0.12$ \text{\AA}. Thus, after the dimerization only the $d_{xy}$-type of orbital ground state exists in the system. Taking into account that $Z_{eff}$ in the crystal environment is lower than its value for the free ion this state can be realized even at a lower critical value of the dimerization shift $\delta c$. In the lower panel of Fig. 7 we illustrate the calculated electron density distribution for the orbital ground state in the SP phase. The appearance of the new $d_{xy}$ orbital order can explain an unusual expansion of the $a$ and $b$ lattice parameters with decreasing temperatures which is most pronounced between $T^*$ and $T_{SP}$ and becomes static below $T_{SP}$.~\cite{law-11,law-thesis}

Taking the orbital reordering process into account, we can qualitatively explain the discrepancy between the intrachain exchange coupling extracted from magnetic Raman scattering in the SP phase and from magnetic susceptibility in the high temperature phase. In accordance with Goodenough-Kanamori rules, the orbital alternation from $e_{g}$-type into $d_{xy}$-type (shown in Fig. 7) drastically decreases the magnitude of intrachain exchange interaction in the SP-phase compared to the high temperature phase. It is also noteworthy that the topology of the electron lobes in the isotropic chain phase would support a direct exchange between Ti-Ti ions instead of a superexchange, as the lobes reach from metal to metal, past neighboring oxygens. This question is highly debated, owing to the anomalously large value of intrachain exchange interaction in TiO$X$ ($X$ = Cl, Br) compounds.~\cite{Shaz}

The symmetry of the first excited crystal field level changes in the opposite way than the symmetry of the ground state level for all cases considered above. For the $e_{g}$, $d_{z^{2}-x^{2}}$-type ground state the first excited level is of $t_{2g}$, $d_{xy}$-type symmetry and vice versa. In the following we will discuss possible reasons for this unusual reverse order ($e_{g}$ instead of $t_{2g}$ symmetry) of the lowest crystal field level in the TiO$_6$ octahedra as well as the source of orbital instability. In TiPO$_4$ the PO$_4$ tetrahedron is the most rigid unit with a high covalence of the P-O bonds. Therefore, the TiO$_6$ octahedra are secondary units with highly distorted ``metal-ligand'' bonds compared to their equilibrium values, which can be estimated from the Shannon radii.

Indeed, a closer look at the high temperature structure reveals that the length of the Ti-O bonds along the $b$-direction is compressed compared to the Ti-O bonds within the $ac$-plane. Therefore, in first approximation the symmetry of the crystal field forces the Ti$^{3+}$ electron density into the $ac$-plane. The lobes of this density distribution should be pointing in between the four oxygen ions of the $ac$-plane (see upper panel of Fig. 7). Thus, we obtain an $e_g$ $d_{z^{2}-x^{2}}$ orbital ground state instead of the expected $t_{2g}$ $d_{xz}$ for an octahedral environment. The strength of crystal field increases as the lattice contracts under lowering the temperature and the influence of oxygen ions situated in the $ac$-plane leads to a reordered orbital ground state into the $ab$-plane (see lower panel of Fig. 7).

The accidental degeneracy of crystal field levels has also some dynamical aspect. Calculations of the adiabatic potential for Ti ions along the $c$-axis (not shown) show that for $Z_{eff}$ larger than a critical value this potential has a minimum at the center of the octahedra. However, under approaching the critical value the depth of the adiabatic potential decreases and it becomes more shallow. For $Z_{eff} < \bar{Z}_{eff}$, the adiabatic potential acquires a double well shape so that the energy minimum is reached at a certain off-centering of the Ti ions. Thus, the accidental degeneracy leads to a Jahn-Teller like instability which is removed by shifting Ti ions along the crystallographic $c$-axis. On the other hand, the magnetic subsystem prevents a cooperative Jahn-Teller distortion with an equidistant distribution of Ti ions. Instead, the magnetoelastic interaction, acting against the Coulomb repulsion, generates Ti dimers. One can conclude that the interplay of the orbital and the magnetic subsystems as well as the orbital instability highly enforces the SP transition, leading to an enhanced $T_{SP}$, much larger than expected from a purely magnetoelastic mechanism. Similarly, the spin gap also increases anomalously as for the closing of the gap one should spend energy against two subsystems, $i$) the magnetic exchange coupling and $ii$) the reordered orbital ground state.

The mode around 3430 cm$^{-1}$ depicted in Fig. 8 is too high in energy for phonon or magnetic scattering processes. A good candidate could be a crystal field excitation. From theoretical calculations, a crystal field excitation energy of 3400 cm$^{-1}$ for the transition from the Ti$^{3+}$ $d_{xy}$ to the $d_{3z^2-x^2}$ level is found in the SP phase at reasonable $Z_{eff}$ values and Ti off-centering shifts (see lower panel of Fig. 6). Furthermore, its energy is of the same order of magnitude as the lowest titanium crystal field onsite excitation observed in TiOCl by resonant inelastic X-Ray scattering.~\cite{Glawion} However, a strong modification of this signal with temperature, as shown in Fig. 8, is unexpected. On the other hand, a jump in frequency under a first order phase transiton at $T_{SP}$ is characteristic for a one-particle scattering process. Note, that such a sensitivity to the lattice distortions can not be observed for two-magnon scattering or for any composite, two-particle processes as they are integrated over the Brillouin zone.

It is also noteworthy that this mode is only observed in $cc$ polarization, i.e. with the light polarization parallel to the spin chains. No signal could be detected in either crossed or perpendicular polarization. Additionally, a change in laser line from $\lambda = 532$ nm to 561 nm had no influence on the mode's energy or the line shape, proving its intrinsic nature in a Raman scattering process. Besides the strong mode at 3430 cm$^{-1}$, weaker shoulders are observed above 3540 cm$^{-1}$, which mimic the temperature and polarization behavior of the former mode. The energy difference between the first shoulder and the main mode corresponds to typical vibrational energies of Ti ions.~\cite{Lopez-Moreno} This points toward very strong coupling between the crystal field excitation and Ti phonons. Respectively, a two-particle process consisting of a phononic and a crystal field excitation can account well for the observed maximum, i.e. the simultaneous scattering with $-k$ to a flat dispersion regime for one excitation and with $+k$ of the same magnitude to a flat dispersion regime of a second excitation. Such a scattering process is schematically sketched in the inset of Fig. 8.

All of these observations support the scenario of a SP transition with a strong amplification by an orbital instability. Another noteworthy anomaly is the large increase of linewidth with temperature of the mode around 3430 cm$^{-1}$ above $T_{SP}$, i.e., when the system enters the incommensurate phase. As illustrated in the lower panel of Fig. 6 the energy of the first crystal field excitation has a very steep slope as function of $Z_{eff}$, so that minute changes of the covalency degree lead to a large variation of the energy. Furthermore, the lower panel of Fig. 6 shows that a much stronger energy variation takes place for modest off-centering shifts of the Ti ions. Note, that in between $T_{SP}$ and $T^*$ the adiabatic potential changes its shape along the chain direction from a two-minimum to a one-minimum form. Thus, the thermal off-centering fluctuations are much larger in the incommensurate phase. Consequently, at $T > T_{SP}$ the dynamical (i.e. thermal) variation of the crystal field excitation energy that effectively manifests itself as an increase in linewidth, is much larger than the natural width. This dynamical width should follow the amplitude behavior of the thermal off-centering fluctuations with temperature proportional to $\sqrt{T-T_{SP}}$ as is observed from experiment (see the orange fit curve to the linewidth in the right panel of Fig. 8).

The orbital instability due to an off-centering dimerization shift of the Ti ions implies that the TiO$_6$ octahedral complex can be easily polarized along the $c$-direction. Therefore, not only the first crystal field excitation but also the Ti phonon modes which oscillate along the $c$-axis should gain in Raman scattering intensity in $cc$-polarization. Based on this we attribute two modes at 132 cm$^{-1}$ and at 166 cm$^{-1}$ to the phonon modes with predominantly contribution from Ti off-centering vibrations. These phonon modes should evidence strong coupling to the first crystal field excitation which is indeed observed as exciton-phonon shoulder around 3540 cm$^{-1}$. Also these phonon modes strongly modulate the magnetic intrachain exchange that is seen as characteristic Fano-like line shape. The asymmetry decreases with temperature, i.e., with the decrease of the modulation amplitude.

In conclusion, we suggest that the Ti orbital degrees of freedom play an essential role in the dimerization process in TiPO$_4$ as their energies are comparable to the energy of the magnetic and lattice subsystems in a certain temperature regime. Furthermore, the orbital instability highly enforces the SP transition by amplifying the value of the spin gap. This situation is completely different in the more conventional SP compound CuGeO$_3$. Comparing the observed effects to TiOCl and TiOBr, it is evident that with respect to the reduced gap ratio as well as to the magnitude of the phonon anomalies the two latter compounds behave even more exotically. The clear identification of mixed orbital-lattice-spin excitations in the rather one-dimensional TiPO$_4$ would then allow to assign the previously identified magnetic scattering in TiOCl and TiOBr at least to a certain degree to mixed excitation processes. To some extent this clears up the previously unsolved discrepancy of their similar energies despite reasonably different exchange coupling constants.

\section{Summary}

We have presented a comprehensive Raman scattering study of the 1D spin chain system TiPO$_4$. This system undergoes a complex dimerization process from a uniform to an incommensurate to a commensurate phase. In the incommensurate phase unusual lattice dynamics is observed which can be ascribed to strong spin-orbital-phonon coupling. The magnetic scattering that shows a pronounced renormalization within this process can be separated into a bosonic low and a fermionic high energy part. From the low energy regime we extract the size of the rather large spin gap $\Delta \approx 330$ K, which points to a contribution of orbital degrees of freedom to the energy gain in the dimerized phase. In contrast to other SP systems, a significant amount of undimerized spins remain in TiPO$_4$ below the SP transition temperature, as evidenced by a considerable spectral weight below the spin gap energy. Furthermore, a mode at high energies with a distinct temperature and polarization dependence is attributed to a crystal field excitation of mixed orbital and lattice nature. We find a huge difference between the value of intrachain exchange interaction extracted from the cut-off magnetic Raman scattering in the SP phase for $T < T_{SP}$ and from the magnetic susceptibility measured for $T > T_{SP}$. We attribute this difference to a different orbital order in TiPO$_4$ at high and at low temperatures. Our modified crystal field cluster calculations support this conclusion. We show that orbital instability develops in the incommensurate phase. The accidental degeneracy of crystal field levels and orbital instability is removed by the SP dimerization process at which strong magnetoelastic coupling acts against the Coulomb repulsion. We uncover an exotic dimerization process different from other SP materials in which the spin gap energy gain is formed by magnetoelastic and orbital mechanisms. Our study highlights the peculiar nature of TiPO$_4$, which reaches far beyond a conventional SP picture.

\begin{acknowledgments}
We acknowledge support from DFG, B-IGSM and the NTH School \textit{Contacts in Nanosystems} and thank K. Schnettler for important help. YGP and KVL were supported in part by NAS of Ukraine Grants No. 27-02-12 and No. 91/13-H. KYC acknowledges financial support from the Alexander-von-Humboldt Foundation and Korea NRF Grants No. 2009-0093817 and No. 2012-046138.
\end{acknowledgments}


\begin{thebibliography}{9}
\bibitem{remark}
The description of the $t_{2g}$ and $e_g$ functions is given in the crystal's coordinate system. In spite of this nonstandard setting the lobes of our ground state orbitals are always directed in between ligands.
\end{thebibliography}
\end{document}